\documentstyle[aps,prl,epsf,preprint]{revtex}
\begin{document}
\title{Spin-Peierls states of quantum antiferromagnets\\ on the $Ca V_4 O_9$
lattice}
\author{Subir Sachdev and N. Read}
\address{Department of Physics, P.O. Box 208120, Yale University,
New Haven, CT 06520-8120}
\date{April 17, 1996}
\maketitle
\begin{abstract}
We discuss the quantum paramagnetic phases of Heisenberg
antiferromagnets on the 1/5-depleted square lattice found in
$Ca V_4 O_9$. The possible phases  of the
quantum dimer model on this lattice are obtained by a mapping
to a quantum-mechanical height model. In addition to the ``decoupled'' phases
found earlier, we find a possible intermediate spin-Peierls phase with
spontaneously-broken lattice symmetry. Experimental signatures of the
different quantum paramagnetic phases are discussed.
\end{abstract}
\pacs{PACS numbers:}
The recent observation of a two-dimensional gapped quantum paramagnet
in $Ca V_4 O_9$~\cite{expt} has stimulated a number of theoretical
studies~\cite{others,ueda,gelfand}
of the spin $S=1/2$ antiferromagnetic Heisenberg model with Hamiltonian
$H = \sum_{<ij>} J_{ij} \hat{S}_i \cdot \hat{S}_j$, and
$\hat{S}_i$ a spin $S$ operator on each site $i$
of the $1/5$-diluted square
lattice (Fig~1) found in this material. Of particular interest is
 the nearest-neighbor model, which has two exchange constants $J_A$, $J_{BC}$
(see Fig~1).
 Theoretically, one possibility for
the ground state is an ordinary N\'{e}el state in which the spins are ordered
in opposite directions on the two sublattices. Alternatively, for $J_A \gg
J_{BC}$
or  $J_{BC} \gg J_A$,
the ground state is paramagnetic and its excitations are
well understood; the lattice can be divided into
essentially decoupled pairs ($J_{BC} \gg J_A$) or
quadruplets ($J_{A} \gg J_{BC}$) of spins.
However, the real material is near neither
limit~\cite{gelfand}, and if the ground state is not N\'{e}el ordered, then
there are a number of experimentally important questions about the resulting
quantum paramagnet state. Among them are: does the ground state
always have the full symmetry of the underlying lattice, as do the two states
mentioned above, or can this symmetry be spontaneously broken? If the latter
is the case, what is the nature of the excitations above the gap---does the
system have unbound spin-1/2 spinons, or are they permanently bound
into excitations with integer spin? Here, we will address these
questions using the framework of earlier work on quantum paramagnet states of
two-dimensional antiferromagnets~\cite{early0,early1,early2}.
We will find that while some of the physics is similar to
that studied earlier on other bipartite lattices (the square and the
honeycomb), the 1/5-diluted square lattice also displays some interesting new
phenomena.

Earlier work considered both unfrustrated~\cite{early0,early1} and
frustrated~\cite{early2} Heisenberg antiferromagnetic Hamiltonians
on the square and honeycomb
lattices. Based on various  methods, a general theory of the quantum paramagnet
phases of such systems  was obtained, and also applied to the triangular and
kagom\'{e}  lattices~\cite{subirkag}. The results are:
(i) on bipartite lattices, with an unfrustrated Hamiltonian, the spinons are
always confined to form integer-spin excitations, and the ground state must
exhibit some form of spin-Peierls ordering that breaks the symmetry of the
lattice, except possibly in the case when $2S$ is a multiple of the
coordination number $z$ of the lattice, when a nondegenerate state with the
full lattice symmetry that has only integer-spin excitations is
possible~\cite{aklt}; (ii) for frustrated systems, such as those with
next-neighbor couplings on bipartite lattices~\cite{early2}, or on
non-bipartite lattices~\cite{subirkag}, a ground state without spin-Peierls
order can occur regardless of the value of $2S$ (mod $z$), with unconfined
spinon excitations, which behave as bosons~\cite{rc}. All these states have
an energy gap for local excitations, except at the second-order
zero-temperature transitions that occur between some of them.

The earlier results for bipartite lattices use the equivalence under lattice
symmetries of all the links (and thus, nearest-neighbor interactions) of the
lattices concerned. This property does not hold for the 1/5-diluted square
lattice, and for the spin-1/2 antiferromagnet there are limiting cases
$J_A \gg J_{BC}$, and $J_{BC} \gg J_A$, in which it is evident that the
ground state is invariant under all the symmetries of the lattice. However,
these states fit into the general picture if we remark that, for non-Bravais
lattices, it is clearly natural to consider the spin per unit cell as the
analog of the spin per site on a Bravais lattice. For the present case, our
lattice has a unit cell of 4 sites (unlike the statement by Ueda
{\it et al.}~\cite{ueda}) and we have possible total spin values of 0, 1,
or 2 on a lattice with square symmetry, so a nondegenerate ground state along
the lines of Ref.\ \cite{aklt} is possible for the spin 2 case; if the sites
of the square lattice are taken to be the $A$ plaquettes, this can be
interpreted as the ground state of the decoupled pair limit ($J_{BC}\gg J_A$).
The decoupled quadruplet limit ($J_A \gg J_{BC}$) is obviously the case of
spin 0 on each $A$ plaquette. On the other hand, when the interactions are
more nearly equal, a full analysis that treats each site as distinct may be
more appropriate. We will show here that the limiting results, and a
spin-Peierls phase, emerge in a unified way by a careful use of our earlier
methods.

Among the various approaches~\cite{early0,early1,zheng,qd2,jalabert} to the
unfrustrated case that lead to the same effective action (all of which can be
applied here), one of the most
appealing is the quantum dimer (QD) model~\cite{qd1,early0}. This model
represents the low-energy spin-singlet states by forming singlet ``valence
bonds'' out of two spins of 1/2 on nearest-neighbor sites. Spin $S$ at a site
can be represented by the symmetrized product of $2S$ spins of 1/2, and hence
there must be $2S$ bonds ending at each site. Here, however, we will
consider $S=1/2$ (the properties for other $S$ are quite similar)
and in this case each basis state of the
QD model is associated with a close-packed dimer covering of the
lattice under consideration, each dimer representing a valence bond. The
effective Hamiltonian in this space consists of local diagonal potential
energy terms, and off-diagonal terms which produce local rearrangements of
the dimer packing.

It has been shown~\cite{qd2,early1,levitov} that the imaginary time path
integral of the QD model can be recast as a ``height'' or ``roughening'' model
in space and imaginary time. The mapping relies on a one-to-one mapping
between every state of the QD model and the equivalence classes of a
configuration of heights, $h_a$ (on the sites, $a$, of the dual lattice
(Fig~1)), under $h_a \rightarrow h_a + 1$. For the 1/5-depleted square lattice
we have
\begin{equation}
h_a = n_a + \zeta_a
\label{hval}
\end{equation}
where $n_a$ is an integer which fluctuates from site to site, while
$\zeta_a$ is a fixed fractional offset: $\zeta_a = 0,\alpha,-\alpha$
for $a \in A,B,C$ respectively, with $1/4<\alpha<1/2$ a fixed real constant.
All configurations of $h_a$ which satisfy the constraint
\begin{equation}
|h_a - h_b| < 1 \qquad \mbox{for every nearest-neighbor pair $a,b$}
\label{constraints}
\end{equation}
are permitted (two sites are nearest neighbors if their plaquettes on the
direct lattice share a link). To map the $h_a$ configurations onto states of
the QD model, examine the value of $h_a - h_b$ for
every pair of nearest neighbors, and if $|h_a - h_b| > 1/2$, only then place a
dimer on the link of the direct lattice shared by the plaquettes around $a$ and
$b$. It can be shown that our choices for the offsets $\zeta_a$ ensure
that there is exactly one dimer terminating at each site of the direct
lattice.

It is helpful to examine the heights associated with some regular dimer
coverings of the 1/5-depleted square lattice. The covering (I) (``flat'')
in Fig 2 corresponds to
$h_A = 0$,
$h_B =
\alpha$, and $h_C = - \alpha$; notice that this is the unique covering which is
invariant under all the symmetries of the underlying lattice. The covering
(II) (``spin-Peierls'') corresponds to $h_A = 0$, $h_B =\alpha$,
and $h_C = 1 - \alpha$; this covering has a partner, under the symmetry
operations of the lattice that interchange $B$ and $C$,
which has $h_A = 1$, $h_B = \alpha$, $h_C = 1 - \alpha$.

The path integral of the quantum mechanical height model
involves summing over spacetime-dependent configurations of the heights,
subject to the constraints, with an action related to the effective Hamiltonian.
We will next write down a phenomenological effective action, ${\cal S}$, that
correctly describes the height model at long length and time scales; the same
method can also be used for the classical dimer packing problem simply by making
all the fields time-independent. We begin by using Poisson summation to promote
the field $h_a$, which can only take the discrete set of values
(\ref{hval}), to a
field $\chi_a$ which can take all real values (the
$\zeta_a$ offsets in (\ref{hval}) require only a slight modification of the
usual method~\cite{early1,qd2}). The constraints (\ref{constraints}) are 
``softened'' by adding
appropriate potential energy terms in the action; we will argue later that
this softening cannot modify the main qualitative features of the results.
We have:
\begin{equation}
{\cal S} = \int d \tau \left[
\sum_{<ab>} K_{ab} (\chi_a - \chi_b )^2 +
\sum_{a} \left\{ K_{\tau a} \left( \partial_{\tau} \chi_a \right)^2
- y_a \cos(2 \pi (\chi_a - \zeta_a)) \right\} + \ldots \right].
\label{action1}
\end{equation}
The terms shown are only representative; any term obeying the
``translational'' symmetry of the height model,  $\chi_a \rightarrow
\chi_a + 1$, is permitted.
The coupling constants, $K_{ab}$, $K_{\tau a}$
and
$y_a$, depend only on the  sublattice label of the sites, $a$, $b$, and the
symmetry of the lattice  requires that $K_{AB} = K_{AC}$, $y_C = y_B$ etc.

We now change variables from the three sublattice fields
$\chi_A$, $\chi_B$, $\chi_C$ to the fields $\chi_1$, $\chi_2$, $\chi_3$ which
are related by
$\chi_A = \chi_1 + \chi_2$, $\chi_B = \chi_1 - \chi_2 + \chi_3$,
$\chi_C = \chi_1 - \chi_2 - \chi_3$.
Now the translational symmetry of the height model affects only $\chi_1$,
$\chi_1 \rightarrow \chi_1 + 1$, and this will lead to important
simplifications. It is useful to note here the values of $\chi_1$, $\chi_2$,
$\chi_3$ for the regular dimer coverings considered earlier.
The state (I) (``flat'') in Fig 2 now corresponds to
$\chi_1 = 0$, $\chi_2 = 0$, $\chi_3 = \alpha$.
The two spin-Peierls states ((II)) have $\chi_1 =  1/4$, $\chi_2 = - 1/4$,
$\chi_3 = \alpha - 1/2$, and $\chi_1 = 3/4$, $\chi_2 = 1/4$, $\chi_3
= \alpha-1/2$. These values suggest that $\chi_2$ plays the role of the
spin-Peierls order parameter, and that a non-zero mean-value of $\chi_2$
implies a spontaneous breaking of the $B \leftrightarrow C$ interchange
symmetry.

The invariance of
$\chi_2$ and $\chi_3$ under translations in height space implies that
``mass'' terms like $m_2^2 \chi_2^2 + m_3^2 \chi_3^2$ can, and do, appear
in the action ${\cal S}$. It is therefore safe to integrate out $\chi_2$ and
$\chi_3$, which obey
\begin{equation}
\chi_2 \sim -\sin (2 \pi \chi_1) ~~~,~~~\chi_3 \sim \cos (2 \pi \chi_1),
\label{corres}
\end{equation}
at their saddle points, and obtain an effective action for the single scalar
field
$\chi_1$. We take  the spatial continuum limit for $\chi_1$ and obtain
\begin{equation}
{\cal S}_1 = \int d \tau d^2 x \left[ K ( \nabla \chi_1 )^2
+ K_{\tau} (\partial_{\tau} \chi_2 )^2 - y_1 \cos(2\pi\chi_1)
- y_2 \cos (4 \pi \chi_1 ) - \ldots \right]
\label{s1}
\end{equation}
Additional terms with more gradients of $\chi_1$, or cosines of higher
integral multiples of $2 \pi \chi_1$ are also present.
Note that, while, for the square (honeycomb) lattices considered
earlier~\cite{early1}, lattice symmetries require that the effective
potential for $\chi_1$ contain only cosines of integral multiples of
$8 \pi \chi_1$ ($6 \pi \chi_1$) (a fact intimately linked
to the ubiquity of spin-Peierls order),
here, lattice symmetries only impose the
requirement that the effective potential be even in $\chi_1$, and we will see
that phases without spin-Peierls order are also possible.

As ${\cal S}_1$ describes a three-dimensional interface, the interface must
always be flat, and the symmetry $\chi_1 \rightarrow \chi_1 + 1$
is spontaneously broken, even if all $y_1$, $y_2$, \ldots, are
zero. The value of $\langle \chi_1 \rangle$, along with (\ref{corres}),
identify the state of the interface, and also of the QD model. A
simple picture of the phases and phase transitions is obtained
by truncating the on-site periodic potential to the two cosine terms
explicitly displayed in (\ref{s1}), and minimizing the energy; more general
potentials have the same qualitative features. The results are shown in Fig 2.
There are three possible phases (modulo global translations of
$\chi_1$ by integers):\\ (I) $\langle \chi_1 \rangle = 0$: This is the ``flat''
phase. There is no broken lattice symmetry, which is consistent with $\langle
\chi_2\rangle = 0$. \\
(II) $0 < \langle \chi_1 \rangle < 1/2$: This state has spin-Peierls
order and has spontaneously broken the $B \leftrightarrow C$ interchange
symmetry because $\langle \chi_2 \rangle \neq 0$. The partner state is obtained
by $\chi_1 \rightarrow 1-\chi_1$, $\chi_2 \rightarrow -\chi_2$,
$\chi_3 \rightarrow \chi_3$.\\
(III) $\langle \chi_1 \rangle = 1/2$: this phase is similar to the
``disordered flat'' phase found in earlier work on two-dimensional
interfaces~\cite{dof}, so we will use that terminology~\cite{square}.
The interface is flat on large scales, but  each $h_A$ fluctuates between
two neighboring values with equal probability for each, and correlations
between $h_A$'s at different sites decay exponentially with separation.
In the QD language, each $A$ plaquette has two dimers which resonate between
the two possible orientations, with correlations between the orientation
of two plaquettes decaying exponentially. There is again no broken symmetry as
$\langle \chi_2 \rangle = 0$.
Both the phases (I) and (III) are invariant under all lattice
symmetries.
They are nevertheless distinct states~\cite{ueda}
which cannot be continuously
connected. There is a non-trivial ``topological order'', measured by the
mean height, which distinguishes them.

It is clear that phases (I) and (III) correspond to the $J_{BC} \gg J_A$
and $J_A \gg J_{BC}$ limits~\cite{ueda} of the
underlying antiferromagnet, respectively. Upon interpolating between these
limits, we move along a section through the phase diagram, and there are
two basic possibilities: ({\em a\/}) There is a direct first order transition
between phases (I) and (III) as occurs in Fig 2 for $y_2 > 0$.
({\em b\/}) The spin Peierls phase ((II)) appears in between (I) and
(III), as is the case in Fig 2 for $y_2 < 0$. A third possibility is,
({\em c\/}), an intermediate phase with N\'{e}el long range order.
This phase clearly lies beyond the scope of the QD model, and we
expect that it can undergo direct second-order transitions to any of the
three quantum paramagnetic phases, though if N\'{e}el and spin-Peierls phases
are both present, we expect them to be adjacent. Finally, we also remark that
the paramagnetic and N\'{e}el phases are stable under addition of not too large,
non-nearest-neighbor, frustrating exchange interactions, although the N\'{e}el
region is expected to shrink in size as frustration increases, as in Ref.\ 
\cite{early2}.

Beyond mean-field theory, fluctuations are expected to be relatively
innocuous. First, as the interface is flat in each phase, the softening of the
constraint (\ref{constraints}) is not expected to have serious consequences;
indeed, imposing (\ref{constraints}) rigidly can only make the interface
flatter. A direct transition between (I) and
(III) (or other flat phases with different mean heights) must always be
first-order when the dimension of spacetime is greater than 2, because of the
spontaneous breakdown of translational symmetry in $\chi_1$ even when the
cosines are absent. In the spin-Peierls phase, the mean height varies
continuously, and we expect a $d=3$ Ising transition to both (I) and (III),
though first order behavior is not ruled out. The Ising-like behavior can
best be understood by expanding the action in powers of
$\chi_1$ about zero (or 1/2) to obtain a Landau-Ginzburg-Wilson $\phi^4$
action, with $\chi_1$ (or $\chi_1-1/2$) playing the role of $\phi$; the
periodicity can be neglected here, since there is a much larger energy barrier
for fluctuations changing $\chi$ by $\pm 1$.

To describe excitations of the antiferromagnet with non-zero spin, we use
earlier results~\cite{early1}, which go beyond the QD model, and show
that a flat interface implies that spinons are confined. For the present model,
this means that all three paramagnetic phases have only integer-spin
excitations.

We now turn to the classical, finite temperature ($T$) dimer model, whose height
representation is the action ${\cal S}_1$ without the time variable.
The phase diagram in ($K$, $y_1$, $y_2$)-space is quite
complex, but can be understood by fairly standard methods, so we will just
summarise the results. The variable $K\sim 1/T$.
At low $T$, a constant $T$ section of the phase diagram near $y_1$, $y_2=0$
resembles Fig.~2; the (I)--(III) transition is still first-order, while the
(I)--(II) and (II)--(III) transitions are expected to be two-dimensional
Ising transitions. For $T$ greater than some critical value, the first-order
transition line becomes second-order in the region adjoining the point where
the three phase boundaries meet. There are then two Kosterlitz-Thouless (KT)
critical points on the (I)--(III) boundary, one where the behavior turns from
first- to second-order, the other where the second-order portion meets the two
Ising phase boundaries. The second-order portion of the (I)--(III) boundary is
a region of continuously-varying exponents; the point $y_1=y_2=0$ lies in this
interval~\cite{salinas}. As $T$ is raised further, a ``rough'' phase,
with logarithmic
correlations of $\chi_1$, appears around $y_1=y_2=0$, separated from (I) and
(III) by a standard roughening transition, which is again of KT
type~\cite{kosterlitz}. While completing
this paper, a paper appeared by Weichman and Prasad~\cite{weichman}, who
analyzed the two-dimensional height model with the two-cosine potential,
finding similar results, but applied to roughening (see the inset to their
Fig.~4). In the literature on classical two-dimensional roughening, a line of
continuously-varying exponents is known as a preroughening
transition~\cite{dof}, and the phase II as
$\theta$-disordered-flat~\cite{weichman,dennijs}.

For the antiferromagnetic Hamiltonian we started with, the
situation is complicated by thermal excitation of nonzero spin excitations, and
the finite $T$ behavior remains unclear to us, though we are confident
that the three phases are separated by phase boundaries at low $T$, and that
the lattice symmetry is broken below a finite $T_c$ in the spin-Peierls region.

An experimental signature of the spin-Peierls phase (II) would be an
accompanying lattice distortion, in which the links on which the spin-spin
correlation is strongest, as shown in Fig.\ 2, would be shortened relative
to the others, reducing the symmetry of the lattice. This would set in below
the transition temperature $T_c$ set by the antiferromagnetic interactions.
It is not yet clear to us how the other phases (I) and (III) can be
experimentally distinguished from one another, although in principle a jump
in the lattice constants should be observable when the first-order transition
at low $T$ is crossed. In all three phases, the elementary spin-carrying
excitations have integer spin, and spin $1$ is most likely. This should
have a clear signature in polarized neutron scattering: the dynamic structure
factor $S(k,\omega)$ ($k$ and $\omega$ are the transferred momentum and
frequency) should have a quasiparticle delta function
$\sim\delta(\omega-\epsilon(k))$ at $T=0$, where $\epsilon(k)$ is the
dispersion relation of the spin 1 quasiparticle. This signature is analogous
to those of $S=1$ Haldane gap antiferromagnets in $d=1$, and is common to
all phases in which spinons are confined, of which a survey was given at
the beginning of this paper.

To conclude, our main prediction is the appearance of the spin-Peierls phase,
as shown in the phase diagram in Fig.\ 2.

We thank R.R.P. Singh and C.L. Henley for stimulating our interest on this
problem, and especially the latter for pointing out that height formulations of
the QD model
could be extended to
the 1/5-depleted square lattice.
This research was supported by NSF grants nos.\ DMR-92-24290 and DMR-91-57484.

\begin{figure}
\epsfxsize=5.8in
\centerline{\epsffile{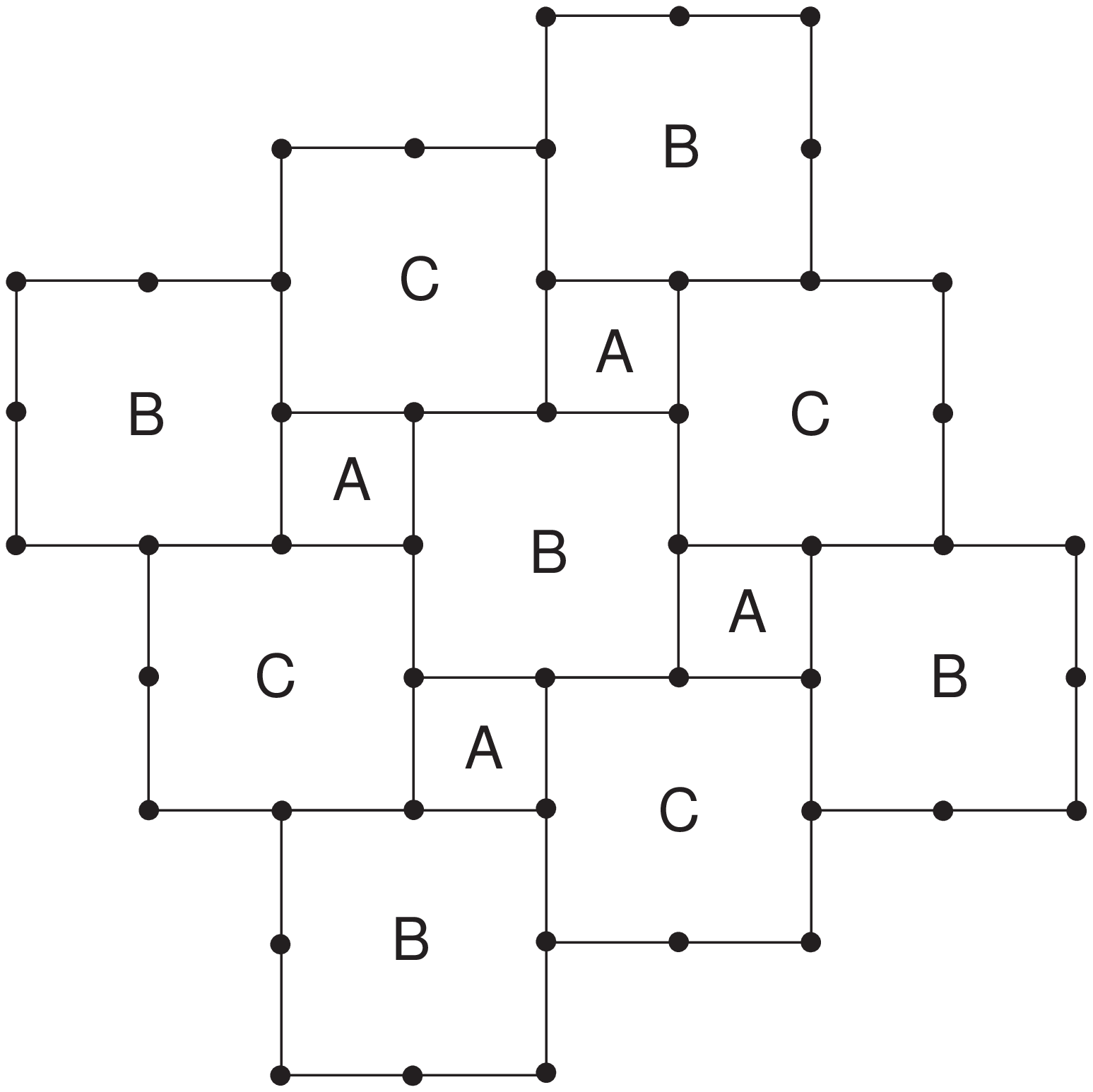}}
\vspace{0.5in}
\caption{The 1/5-depleted square lattice, and the three sublattices ($A$, $B$,
$C$) of its dual lattice. The exchange $J_A$ acts between spins on a link
in a $A$ plaquette, while $J_{BC}$ acts on links shared by the $B$ and $C$
plaquettes. For $J_{BC} \gg J_A$, the ground state is in the
``flat'' phase of Fig.~2, and for $J_A \gg J_{BC}$ it is in the
``disordered flat'' phase.}
\label{f1}
\end{figure}
\begin{figure}
\epsfxsize=5.8in
\centerline{\epsffile{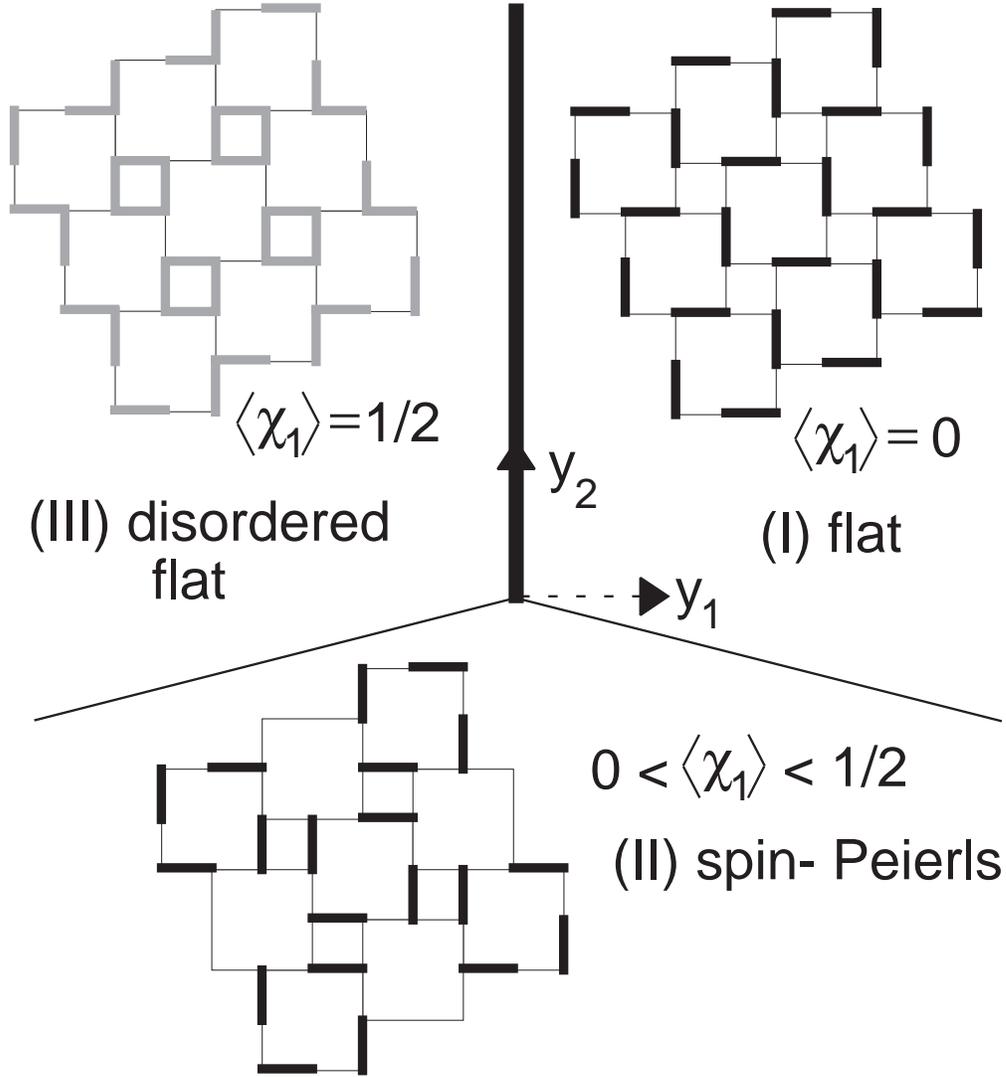}}
\vspace{0.5in}
\caption{Mean-field phase diagram of the model defined by $S_1$
(Eqn (\protect\ref{s1})) with the potential truncated to two cosines.
The thick line is a first order transition, while the thin lines are second
order. Effects of fluctuations are discussed in the text.}
\label{f2}
\end{figure}

\end{document}